\documentclass{ws-procs9x6}

\begin{document}

\title{Implications of the JHF-Kamioka neutrino oscillation experiment}

\author{R.~R. VOLKAS}

\address{School of Physics\\ 
Research Centre for High Energy Physics\\ 
The University of Melbourne\\
Victoria 3010 Australia}

\maketitle

\abstracts{After quickly reviewing the existing evidence for neutrino oscillations,
I summarise the goals and capabilities of the JHF-Kamioka
long baseline superbeam experiment. Theoretical implications
of what this experiment could potentially discover are then discussed.}

\section{Introduction}

Neutrino oscillations arise when there is a mismatch between the neutrino
states produced in weak interaction processes (``weak eigenstates'')
and the Hamiltonian eigenstates (``mass eigenstates''). The three known
weak eigenstates are the familiar $\nu_{e,\mu,\tau}$ flavours. If neutrinos
have non-degenerate masses, then the neutrino mixing matrix $U$
defined through
\begin{equation}
\nu_{\alpha} = \sum_{i=1,2,3} U_{\alpha i} \nu_i,
\label{eq:weakmassgen}
\end{equation}
gives rise to non-trivial effects including oscillations. In this
equation, $\alpha = e,\mu,\tau$ and $\nu_i$ is the state of
definite mass $m_i$. The complex numbers $U_{\alpha i}$
constitute the mixing matrix. 
Additional light neutral fermions usually
known as ``sterile neutrinos $\nu_s$'' may also exist. If so, then
Eq.~(\ref{eq:weakmassgen}) must be generalised in the obvious way.
For the three flavour case, the mixing matrix can be parameterised
in terms of three physical mixing angles and some CP violating 
phases (the precise number of which depends on whether the
neutrino masses are of Dirac or Majorana form). If light sterile
neutrinos exist, then there are additional mass and mixing parameters.

Oscillations arise due to relative phases between the $\nu_i$
induced by time evolution. Considering two flavours only
for simplicity, the mixing pattern
\begin{eqnarray}
\nu_{\alpha} & = & \cos \theta \nu_1 + \sin \theta \nu_2, \nonumber\\
\nu_{\beta} & = & - \sin \theta \nu_1 + \cos \theta \nu_2
\label{eq:2flmixing}
\end{eqnarray}
implies the transition probability
\begin{equation}
P(\nu_{\alpha} \to \nu_{\beta}) = \sin^2 2\theta \, \sin^2 \frac{\Delta m^2 L}{4E},
\end{equation}
after the state which begins life as a $\nu_{\alpha}$ propagates through a distance $L$.
The mixing angle
sets the magnitude of the oscillations, and $\Delta m^2/E$ determines the
oscillation length, where $E$ is energy and $\Delta m^2 = m_2^2 - m_1^2$. 
It is straightforward to generalise this formula to multiflavour
cases. Nature chooses the $\Delta m^2$ and $\theta$ parameters, while
experimentalists have some control over $E$ and $L$. This partial freedom
is utilised in the design of the JHF-Kamioka experiment that is the
focus of this talk.~\cite{loi}

Extremely convincing evidence for the disappearance of muon-neutrinos
has been provided by SuperKamiokande and other experiments through observations
of atmospheric neutrinos.~\cite{atmos}
The upper atmosphere acts as a beam dump for
cosmic rays, with $\nu_{e,\mu}$ and their antiparticles produced as
byproducts. The zenith angle pattern of the contained 
$\mu$-like events reveals a
clear deficit of up-going relative to down-going progenitor $\nu_{\mu}$'s,
while the $e$-like events show no anomalous angular dependence. These data
are consistent with $\nu_{\mu} \to \nu_x$ oscillation, where $x \neq e$.
Doing a detailed fit assuming either $\nu_{\mu} \to \nu_{\tau}$
or $\nu_{\mu} \to \nu_s$ produces allowed regions which
can be roughly described as $\sin^2 2\theta \stackrel{>}{\sim} 0.85$
and $10^{-3} \stackrel{<}{\sim} \Delta m^2/{\rm eV}^2 
\stackrel{<}{\sim} 8 \times 10^{-3}$. Other aspects of the
atmospheric neutrino data show a preference for $\nu_{\mu} \to \nu_{\tau}$
over $\nu_{\mu} \to \nu_s$, the statistical significance of which
has been under dispute.~\cite{dispute} 
JHF-Kamioka and the other long baseline
experiments have been designed to reproduce the atmospheric
neutrino effect in a terrestrial context where the neutrino
source as well as the detector are under experimental control.
Indeed, the pioneering K2K long baseline experiment has already
reported a $\nu_{\mu}$ deficit roughly consistent with the
atmospheric effect.~\cite{k2k}

All solar neutrino experiments have revealed a deficit by
a factor of $2-3$ in the
$\nu_e$ flux relative to standard solar model expectations.~\cite{solar,sno} The
new data from SNO provide strong evidence that solar $\nu_e$'s
oscillate into other active flavours en route to the Earth.~\cite{sno}
The spectrally undistorted nature of the $^8B$ neutrino
flux, when combined with the strong deficit factor, 
limits the oscillation parameter space to $\sin^2 2 \theta
\stackrel{>}{\sim} 0.7$. It is interesting that both the
solar and atmospheric mixing angles are large, quite
unlike their quark analogues. The solar $\Delta m^2$ is constrained
to be at least an order of magnitude smaller than its
atmospheric counterpart.

The LSND experiment has provided fully terrestrial evidence
for $\overline{\nu}_{\mu} \to \overline{\nu}_e$ oscillations,
with a small mixing angle and a relatively large $\Delta m^2$
of about $1$ eV$^2$.~\cite{lsnd} 
This as yet uncorroborated but fascinating
result will soon be checked by MiniBooNE. Following a 
common practice that I do not condone, I will sometimes
``bury my head
in the sand'' \cite{adr} during this talk by assuming that the
LSND anomaly is not due to oscillations. If 
all three anomalies {\it are} due to oscillations, then 
the incommensurate $\Delta m^2$ values imply that at
least one additional flavour, necessarily sterile, must exist.

So, in summary, with head in the sand: the atmospheric
and solar anomalies imply that two out of the three
mixing angles in $U$ are large (and at least the
atmospheric one can even be maximal). These two
angles are usually denoted $\theta_{12}$ and $\theta_{23}$.
The third mixing angle, $\theta_{13}$, is constrained
to be small through neutrino disappearence bounds,
and we have no constraints on the CP violating phase 
$\delta$.\footnote{Note that the additional phases of
the Majorana case do not affect oscillation probabilities.}
Coming clean with LSND forces us to also confront the
possible existence of $\nu_s$ flavours and additional parameters.

In light of the above, the scientific goals of JHF-Kamioka
are well motivated. The main ones are:
\begin{itemize}
\item precision measurement of the atmospheric neutrino
oscillation parameters;
\item discrimination between $\nu_{\mu} \to \nu_{\tau}$
and $\nu_{\mu} \to \nu_s$;
\item search for $\nu_{\mu} \to \nu_e$;
\item search for CP violation in the lepton sector.
\end{itemize}
The first three of these goals can happen during phase 1
of the project, while the fourth will have to wait
for phase 2.

\section{The capabilities of JHF-Kamioka}

The JHF-Kamioka project envisages a high flux, narrow band
$\nu_{\mu}$ or $\overline{\nu}_{\mu}$ beam 
with peak energy in the few-GeV regime being directed
from the Japan Hadron Facility to the Kamioka laboratory
located about 295 km away. Contamination due to $\nu_e$
will be reduced by having a relatively short decay
volume for the muons produced by pion decay. Various types
of beams will be possible, depending on momentum selection
of the parent pions and the choice of beam direction (off axis
or well-directed). The peak energy will be tunable
and the beam spectrum well known, enhancing
sensitivity to oscillation effects. The far detector
in phase 1 will be the existing Super-Kamiokande 50 kt
water \u{C}erenkov detector. Phase 2 envisages an increase
in beam power and the construction of a 1 Mt
water \u{C}erenkov ``Hyper-Kamiokande'' as a second
far detector. Plots of typical beam spectra including
flavour composition are available from the collaboration's
Letter of Intent.~\cite{loi}

\subsection{Precision measurement of ``atmospheric'' oscillation
parameters.} 
 
\begin{figure}[th]
\centerline{\epsfxsize=12cm\epsfbox{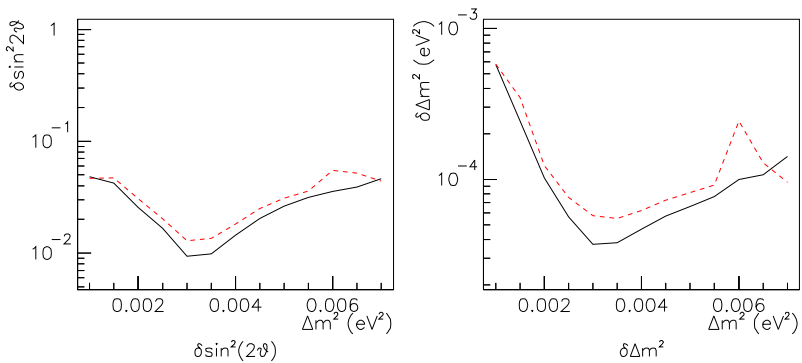}}   
\caption{Sensitivity of the atmospheric oscillation parameters
for the case $\sin^2 2\theta = 0.9$ (dashed line) compared
to the case $\sin^2 2\theta = 1$ (solid line), as a
function of the true $\Delta m^2$. The beam choice has
been optimised for a true $\Delta m^2$ of $3 \times 10^{-3}$
eV$^2$ in this illustration. See the LOI for further details.
\label{fig:atmos}}
\end{figure}

Figure \ref{fig:atmos} depicts the expected
precision for measurements of the oscillation parameters
relevant for solving the atmospheric neutrino 
problem.\footnote{These and all subsequent figures come
from the LOI.~\cite{loi}}
The sensitivity depends on the type of
beam used and on the actual values of the parameters.
Precision measurements down to about $0.01$ in $\sin^2 2\theta$
and ${\rm few}\times 10^{-5}$ in $\Delta m^2/{\rm eV}^2$
are envisaged.

\subsection{Discrimination of $\nu_{\mu} \to \nu_{\tau}$ and
$\nu_{\mu} \to \nu_s$.} 

\begin{figure}[th]
\centerline{\epsfxsize=4cm\epsfbox{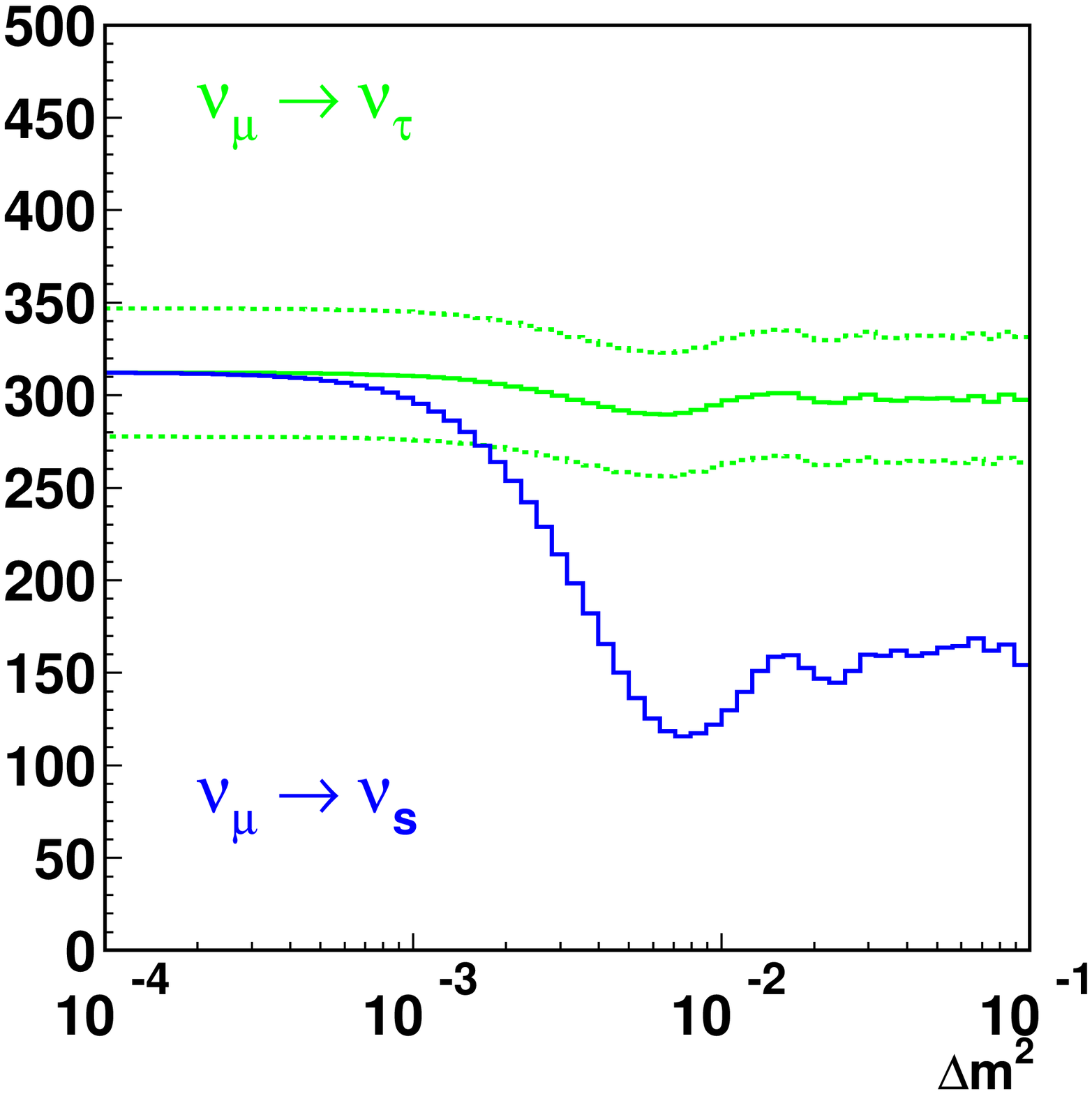}
\epsfxsize=4cm\epsfbox{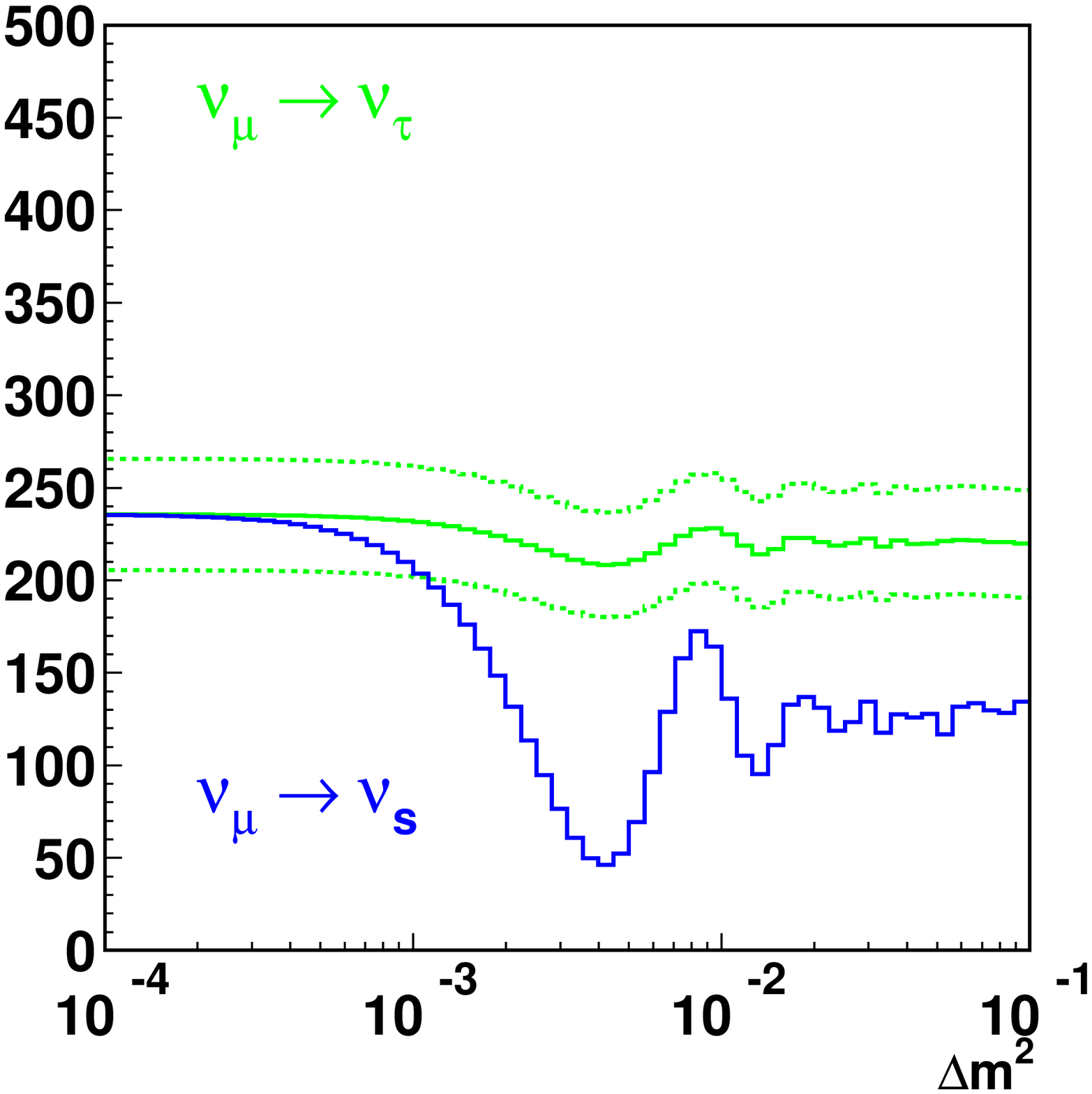}
\epsfxsize=4cm\epsfbox{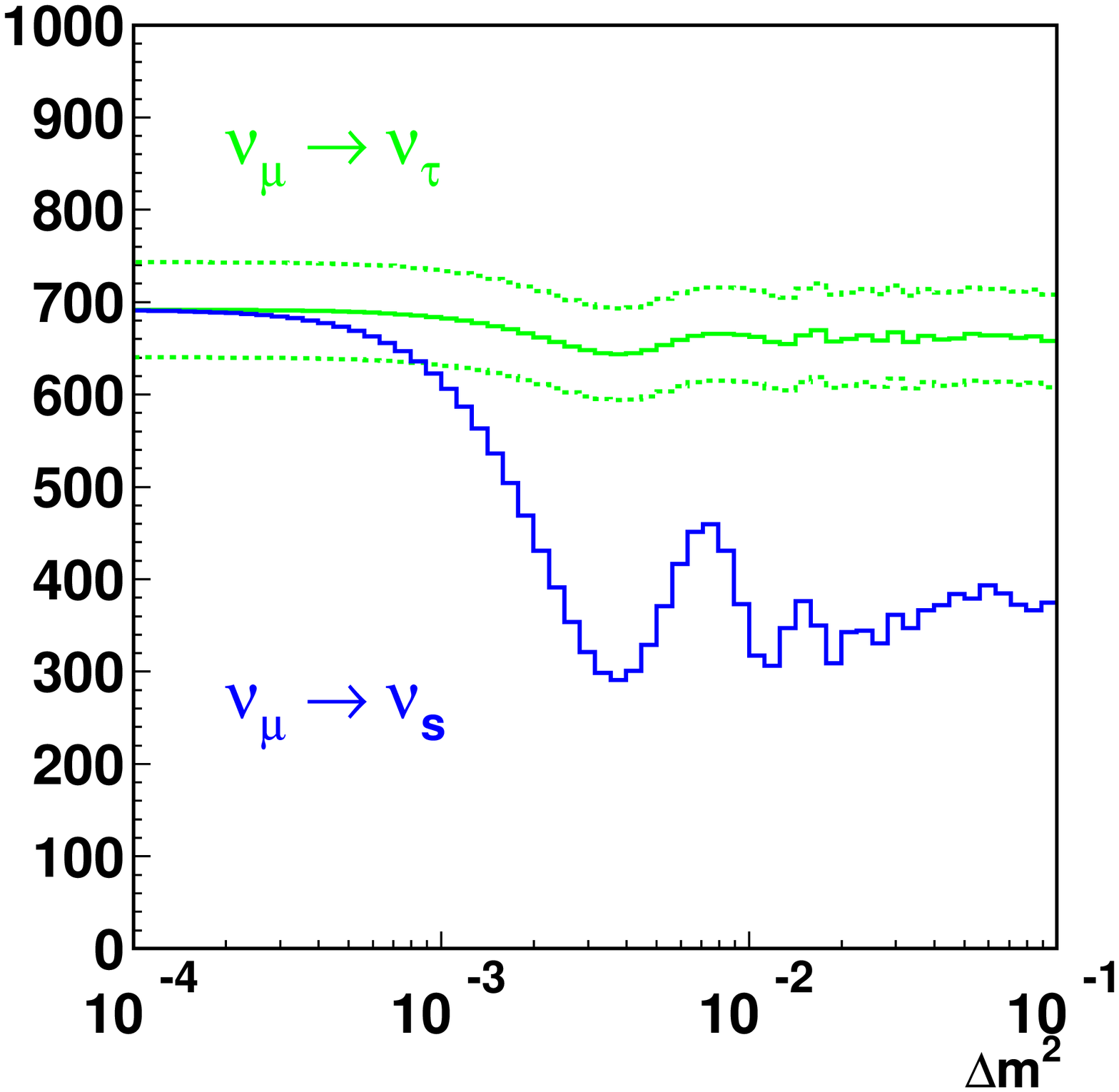}
}   
\caption{Comparison of single $\pi^0$ production
for the $\nu_{\mu} \to \nu_{\tau}$ and
$\nu_{\mu} \to \nu_s$ atmospheric neutrino channels.
The plots assume maximal mixing, with the three
panels corresponding to different possible beams.
Discrimination can be achieved in the $\Delta m^2$
range of interest for resolving the atmospheric
anomaly.
\label{fig:taus}}
\end{figure}

This relies on the observation of
neutral current (NC) induced single pion production in the far detector.
The $\pi^{0}$ events will be the most useful, because of
the relatively clean nature of the $\gamma\gamma$ decay mode.
Figure \ref{fig:taus} compares the expected rates for the
$\nu_{\tau}$ and $\nu_s$ cases, with a clear suppression
evident in the latter for $\Delta m^2 > 1-2 \times 10^{-3}$
eV$^2$.

\subsection{Search for $\nu_{\mu} \to \nu_e$ and CP violation.} 

\begin{figure}[th]
\centerline{\epsfxsize=8cm\epsfbox{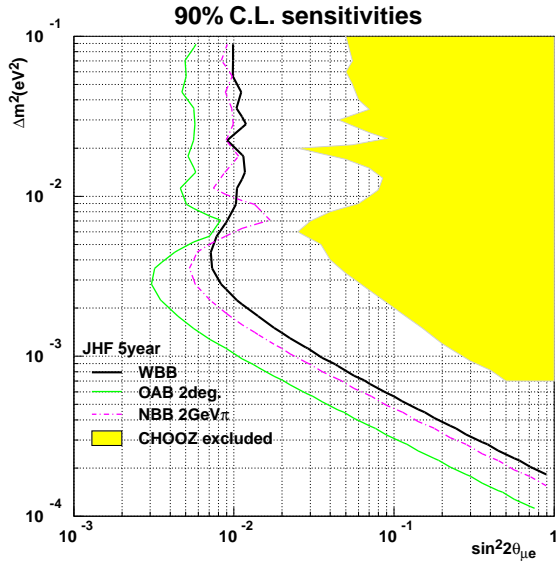}}   
\caption{Expected reach of the
$\nu_{\mu} \to \nu_e$ oscillation search via $\nu_e$
appearance after 5 years of running. The three contours
correspond to different beam choices.
\label{fig:mue}}
\end{figure}

MiniBooNE will
confirm or disconfirm the $\overline{\nu}_{\mu} \to 
\overline{\nu}_e$ interpretation of the LSND anomaly.
Assuming disconfirmation, the existence of such
an oscillation mode again becomes an open question.
In the three neutrino 
picture, the $\nu_{\mu} \to \nu_e$ transition
probability is proportional to the
small parameter $\sin^2 2\theta_{13}$.
Figure \ref{fig:mue} displays the expected
sensitivity for $\nu_e$ appearence in
JHF-Kamioka, with the region
already excluded by CHOOZ and Palo Verde superimposed.~\cite{chooz}
The angle $\theta_{13}$ must
be sufficiently large for CP violation effects to
be observable in oscillation experiments.

%\vspace{-2mm}

\section{Theoretical implications}

I will now briefly discuss possible theoretical ramifications of
the type of information JHF-Kamioka could provide.

\subsection{$\nu_s$ or no $\nu_s$}

Particles, arranged into multiplets, form the raw ingredients
for spontaneously broken gauge theories such as the
standard model. A very basic activity in theoretical
particle physics is to understand how the standard model
Lagrangian might emerge in a effective sense out of a
more fundamental theory. To pursue these studies, we really
need to know what the fundamental low-mass degrees of freedom
are. The possible existence of light sterile neutrinos is
therefore a very interesting loose end from the theoretical
perspective as well as the phenomenological.

One of the famous issues arising from the standard model
is the flavour problem: can the values of the quark and
lepton mass and mixing angle parameters,
and the family structure, be understood
through a standard model extension? Neutrinos
could well provide very important clues, because of the
contrast they provide to the other fermions. Neutrinos
are unusually light, and the large vacuum mixing angles 
required look qualitatively very different to the
small Kobayashi-Maskawa mixing angles of the quark
sector.
But before we can properly reflect on how they might help resolve
(or deepen!) the flavour puzzle, we need to know exactly
how many neutrino-like degrees of freedom exist. The
discovery of sterile neutrinos would be roughly as
important as the discoveries of $c$, $\tau$ and $b$ in
the 1970's.

In the near future, we await results from MiniBooNE.
While this experiment is very important for sterile
neutrino research, it can only provide indirect
evidence for their existence.
Irrespective of what is found by MiniBooNE, the ability
of experiments such as JHF-Kamioka to perform neutral
current measurements and thus potentially
discover sterile neutrinos directly is very welcome. SNO has
of course recently provided strong constraints on
the sterile neutrino component of the solar neutrino
flux.~\cite{sno}

A famous theoretical problem posed by light sterile neutrinos
is: Why {\it are} they light? The most sterile of possible
sterile neutrino candidates are fermions with the
gauge quantum numbers of the vacuum.~\cite{erice} 
Such states obviously
have gauge invariant Majorana mass terms, and there is no 
{\it a priori} reason to expect them to be of similar magnitude
to the active neutrino masses. In fact they can be arbitrarily
large. Mirror symmetry has been proposed to explain both
why apparently sterile states exist and why they are 
light.~\cite{melb,bm}
If the mirror matter idea is correct, then sterile neutrinos
would be just the tip of iceberg, because mirror partners
would be expected for all known particles. The ramifications
of this would obviously be enormous.

\subsection{Precision measurements of the atmospheric parameters.}

As well as performing a degree of freedom audit, we of course
also need as much information as possible on the precise values
of mass and mixing angle parameters. We can dream that one day
a predictive theory for flavour will emerge, and an
important test will be a direct comparison of those predictions
with measured neutrino parameters. In the meantime, we should
try to at least correlate {\it aspects} of the neutrino flavour
problem with new theoretical principles.

The existence of large neutrino mixing angles is thought
provoking. The mirror symmetry idea allows two-flavour
active-sterile {\it maximal} mixing to be understood on the basis of
a simple theoretical principle. We had hoped that the solar
and atmospheric neutrino problems could be solved in a
unified way through the maximal oscillations of $\nu_e$'s
and $\nu_{\mu}$'s into their respective mirror (sterile) 
partners.~\cite{melb}
Alas, the SNO results appear to have ruled out the
solar neutrino part of this hypothesis (they imply an upper
bound on the $\nu_e$--mirror-$\nu_e$ $\Delta m^2$ parameter).
As discussed above, it is important for experiments
such as JHF-Kamioka to check the claim from
Super-Kamiokande that the atmospheric mode is
predominantly into $\nu_{\tau}$.~\cite{fv2002} 

But there is also the question of the atmospheric mixing
angle: is it maximal or merely large? A
precision measurement of $\sin^2 2\theta$
at the $0.01$ level has the potential to rule out
exact maximal mixing, or point more strongly towards it. 
This is important theoretically,
because exact maximal mixing is a special point in
parameter space. The atmospheric neutrino data
have always preferred true maximal mixing, though
it is possible that the actual value is, say,
$\sin^2 2\theta = 0.93$. If so, then JHF-Kamioka
should be able to rule out maximal mixing to a high
level of statistical significance.

Maximal mixing would point to an underlying new
symmetry of nature. If the mode is $\nu_{\mu} \to \nu_s$,
then mirror
symmetry would be the prime candidate. But we should
in general endeavour to discover new symmetry principles
through neutrino oscillation physics. It is perhaps
useful to categorise such attempts according
to whether the symmetry is exact (e.g.\ the Melbourne
version of mirror symmetry), spontaneously broken
(e.g.\ broken mirror symmetry, horizontal symmetry)
or approximate (e.g.\ $L_e \pm L_{\mu} - L_{\tau}$).
There is historical precedent for the first and
third possibilities (e.g.\ colour and electromagnetic
gauge invariance, and Gell-Mann--Neeman SU(3), 
respectively), while the second awaits discovery
of the Higgs boson.

\subsection{$\nu_{\mu} \to \nu_e$ search and CP violation.}

Let us assume that LSND has not already discovered the
anti-particle version of this oscillation mode. 
Then the connection between
$\theta_{13}$ and the existence of CP violation indicates
that the former in a sense quantifies the extent to which the
neutrino mixing is ``truly three-flavour''. This is
an important part of the flavour puzzle.

Observing CP violation in the lepton sector would allow
comparison with similar effects in the quark sector,
with information about the latter on the rise because
of the $B$-factory experiments. 
The neutrino sector already
displays a difference from the quark sector through
its large mixing angles. How will CP violation compare,
and what implications will that have on theories
of quark-lepton symmetry?

CP violation is of course important in theories of baryogenesis.
While its establishment in neutrino oscillations would
have no direct consequence for baryogenesis, it would
show that matter-antimatter asymmetry is not confined
to strongly interacting particles. Baryogenesis can
proceed through the sphaleron reprocessing of a
lepton asymmetry created, for example, from out-of-equilibrium
and CP violating decays of ``heavy neutral leptons''.~\cite{leptogen}
The latter (hypothetical) 
species are neutrino-like, but very massive (perhaps
they are the heavy gauge singlets needed for the
see-saw mechanism). Unfortunately, the CP violating
parameters in the heavy neutral fermion sector need
not be related to those in the light neutral fermion
(i.e.\ neutrino) sector. While these interconnections
are not mandatory, one can hope for relations
within specific and predictive standard model extensions.
We are a long way from having such a theory, but
all the experimental information we can get will help.
Switching perspective, plausible theoretical
proposals for connecting the neutrino
sector parameters to baryogenesis would be welcomed
by experimentalists as a spur to their leptonic
ambitions. The possible existence of Majorana phases
in addition to Dirac phase(s) is an important
consideration.

\section{Conclusion.}

Long baseline superbeam experiments such as JHF-Kamioka
promise to supply very important new information about
the neutrino sector, from precision measurements of
parameters through to possible discovery of sterile
neutrinos and/or CP violation. These are of great importance
in the quest to understand the flavour problem.

While not discussed fully in this talk, 
these results will be of great relevance
for astrophysics and cosmology as well as for particle
physics, especially if light sterile neutrinos are 
discovered.~\cite{cosmo}
The discovery of leptonic CP violation would also be 
(indirectly) important for the baryogenesis puzzle.

\section*{Acknowledgments}
I would like to thank Tony Thomas for inviting me to this
workshop and to the
Special Research Centre for the Subatomic Structure of Matter
at the University of Adelaide for partial
financial support. This work was also partially supported 
by the University of Melbourne. I would also like to 
thank the participants in the neutrino stream at the 
recent WIN meeting in Christchurch for stimulating some of the
thoughts expressed during this talk.


\begin{thebibliography}{99}

\bibitem{loi} Y. Itow et al., hep-ex/0106019.

\bibitem{atmos} Super-Kamiokande Collaboration,
Y. Fukuda {\it et al.}, {\it Phys.
Rev. Lett.} {\bf 82}, 1562 (1998); {\it Phys. Lett.} {\bf B436}, 33 (1998);
{\it Phys. Lett.} {\bf B433}, 9 (1998);
Soudan 2 Collaboration,
W. W. Allison {\it et al.}, 
{\it Phys. Lett.} {\bf B449}, 137 (1999).

\bibitem{k2k}
K2K Collaboration, S. H. Ahn {\it et al.,} {\it Phys. Lett.} {\bf B511},
178 (2001).

\bibitem{dispute}
Super-Kamiokande Collaboration,
S. Fukuda {\it et al.},
{\it Phys. Rev. Lett.} {\bf 85}, 3999 (2000);
R. Foot, {\it Phys. Lett.} {\bf B496}, 169 (2000).

\bibitem{solar}
Homestake Collaboration, B. T. Cleveland {\it et al.}, {\it Astrophys. J.}
{\bf 496}, 505 (1998); Kamiokande Collaboration, Y. Fukuda {\it et al.},
{\it Phys. Rev. Lett.} {\bf 77}, 1683 (1996); Super-Kamiokande
Collaboration, {\it Phys. Rev. Lett.} {\bf 86}, 5651 (2001);
Sage Collaboration, J. N.
Abdurashitov,
{\it et al.}, {\it Phys. Rev. Lett.} {\bf 83}, 4686 (1999);
Gallex Collaboration, W. Hampel {\it et al.}, {\it Phys. Lett.} {\bf B447},
127 (1999); GNO Collaboration, M. Altann {\it et al.}, {\it Phys. Lett.}
{\bf B490}, 16 (2000).

\bibitem{sno}
SNO Collaboration,
Q. R. Ahmad {\it et al.}, nucl-ex/0204008; nucl-ex/0204009;
{\it Phys. Rev. Lett.} {\bf 87}, 071301 (2001).

\bibitem{lsnd}
LSND Collaboration,
C. Athanassapoulos {\it et al.}, {\it Phys. Rev. Lett.} {\bf 81}, 1774 (1998);
{\it Phys. Rev.} {\bf C58}, 2489 (1998).

\bibitem{adr}
As far as I know, this terminology was proposed by A. de Rujula.

\bibitem{chooz}
CHOOZ Collaboration, M. Apollonio {\it et al.,} {\it Phys. Lett.} {\bf B466}
415 (1999); Palo Verde Collaboration, F. Boehm {\it et al.,} {\it Nucl. Phys.
Proc. Suppl.} {\bf 91} 91 (2001).

\bibitem{erice}
For a pedagogical introduction to sterile neutrino theories see
R. R. Volkas, hep-ph/011326, {\it Prog. Part. Nucl. Phys.} (in press).

\bibitem{melb}
R. Foot, H. Lew and R. R. Volkas, {\it Mod. Phys. Lett.} {\bf A7}, 2567 (1992);
R. Foot, {\it Mod. Phys. Lett.} {\bf A9}, 169 (1994);
R. Foot and R. R. Volkas, {\it Phys. Rev.} {\bf D52}, 6595 (1995).

\bibitem{bm}
Z. G. Berezhiani and R. N. Mohapatra, {\it Phys. Rev.} {\bf D52} 6607 (1995).

\bibitem{fv2002}
R. Foot and R. R. Volkas, hep-ph/0204265.

\bibitem{leptogen}
M. Fukugita and T. Yanagida, {\it Phys. Lett.} {\bf B174}, 45 (1986).

\bibitem{cosmo}
For reviews on neutrino cosmology see, M. Prakash, J. M. Lattimer,
R. F. Sawyer and R. R. Volkas, {\it Ann. Rev. Nucl. Part. Sci.} {\bf 51},
295 (2001); A. D. Dolgov, hep-ph/0202122; for recent interesting work on
neutrino oscillations and cosmology see A. D. Dolgov et al.,
hep-ph/0201287; Y. Y. Y. Wong, hep-ph/0203180; K. N. Abazajian,
J. F. Beacom and N. F. Bell, astro-ph/0203442.


\end{thebibliography}
\end{document}